%% file: main.tex
\documentclass[
 aps,
 amsmath,amssymb,
 reprint,
 superscriptaddress,
 nofootinbib,
 floatfix,
 preprintnumbers
]{revtex4-1}
\usepackage[caption=false]{subfig}
\usepackage{bbold}
\usepackage{hyperref}

\usepackage{tikz}

\newtheorem{theorem}{Theorem}
\newtheorem{definition}{Definition}
\newtheorem{lemma}{Lemma}

\usepackage{algorithm}
\usepackage{algpseudocode}

\usepackage{algpseudocode}

\usepackage{comment}

\usepackage{booktabs}
\usepackage{enumitem}

\def\pp{\color{black}}
\def\ora{\color{black}}

\newcommand{\cyan}[1]{{\color{black} {#1}}}

\begin{document}


\title{Reducing circuit depth with qubitwise diagonalization}

\author{Edison M.\ Murairi}
\email{emm712@gwu.edu}
\affiliation{Department of Physics, The George Washington University,
Washington, District of Columbia 20052, USA
}
\author{Michael J.~Cervia}
\email{cervia@umd.edu}
\affiliation{Department of Physics, The George Washington University,
Washington, District of Columbia 20052, USA
}
\affiliation{Department of Physics, University of Maryland,
College Park, Maryland 20742, USA
}

\begin{abstract}
  A variety of quantum algorithms
  employ 
  Pauli operators 
  as a convenient basis for studying the spectrum or evolution of Hamiltonians or measuring multi-body observables.
  One strategy to reduce 
  circuit depth
  in such algorithms involves simultaneous diagonalization of 
  Pauli operators generating 
  unitary evolution operators or observables of interest.
  We propose an algorithm yielding quantum circuits with depths $O(n\log r)$ diagonalizing $n$-qubit operators generated by $r$ Pauli operators.
  Moreover, as our algorithm iteratively diagonalizes all operators on at least one qubit per step, it is well suited to maintain low circuit depth even on hardware with limited qubit connectivity.
  We observe that our algorithm performs favorably in producing 
  quantum circuits diagonalizing randomly generated Hamiltonians as well as 
  molecular Hamiltonians
  with short depths and low two-qubit gate counts.
\vspace{20mm}
\end{abstract}

\maketitle

\section{Introduction}
\label{sec:intro}

Quantum simulations~\cite{ctx21546455140004107, benioff80, feyn81, feyn86, doi:10.1126/science.273.5278.1073, nielsen_chuang_2010, doi:10.1080/00268976.2011.552441} are frequently cited as one of the most promising classes of quantum algorithms to see a practical quantum advantage~\cite{practical_qa}, in an era where analog quantum simulators and universal quantum computers are becoming viable for many-body systems of size ${O}(50)$.
In the case of qubit hardware, it is frequently convenient to decompose one's Hamiltonian into Pauli operators, also known as ``multi-qubit Pauli matrices,'' i.e., Pauli matrices on individual qubits strung together~\cite{nielsen_chuang_2010}.
In fact, this convenience carries over to other algorithms implementing the action of a Hamiltonian on qubits, such as in a quantum Lanczos algorithm suggested by Ref.~\cite{Kirby:2022ncy}.
Moreover, for a yet wider collection of quantum algorithms, efficient measurement schemes must be devised to estimate the value of physical observables with respect to a given prepared state, even where direct simulation of time evolution is not required, such as in variational quantum eigensolvers~\cite{vqe_og} and quantum tomography~\cite{RevModPhys.29.74, PhysRevA.64.052312, paris04, cramer2010, nielsen_chuang_2010}.
Due to the qubitwise nature of measuring prepared quantum states, it is already natural to consider Pauli operators as a basis for measurements of observables in these many algorithms~\cite{RevModPhys.29.74, paris04, nielsen_chuang_2010, vqe_og}.

In practical implementations of such algorithms,
one finds that the number of Pauli operators involved, i.e., in decomposing one's Hamiltonian to be simulated
or physical observables to be measured, may grow polynomially or even exponentially in system size.
This growth can have troublesome consequences for the circuit depth involved in such computations~\cite{Murairi:2022zdg}.
In particular, one finds larger errors will be introduced into computations due to the decoherence of qubits associated with two-qubit gates (with entangling power) in near-term hardware.

As such, one would like to augment these algorithms with strategies to minimize such costs.
One such strategy involves simultaneous diagonalization of commuting Pauli operators~\cite{vandenBerg2020circuitoptimization}, offering a reduction of the problem at hand to only the diagonal Pauli basis (i.e., strings of identity and $Z$ Pauli matrices).
Generally speaking, for measurement schemes on universal quantum computers, one must already perform a sort of diagonalization for each of the Pauli operators generating an observable of interest (since one typically measures in a $Z$ basis as well),
so the suggestion of simultaneous diagonalization~\cite{Crawford2021efficientquantum} is a natural choice to reduce the number of measurements entailed.
However, simultaneous diagonalization for measurements of Pauli operators may entail two-qubit gates that would otherwise be unnecessary in the state preparation or measurements of individual Pauli operators.
Therefore, it is additionally important to minimize the number of such gates involved in simultaneous diagonalization to find the strategy advantageous.
Conveniently, such a strategy can be used in tandem with other procedures to efficiently perform state tomography, such as randomized measurement protocols, e.g., in Ref.~\cite{Bringewatt:2023xxc}.

In what follows, we present a method to exactly and efficiently diagonalize a commuting set of Pauli operators.
To provide context, we give a brief introduction and review to the mathematical language of stabilizer theory typically used to solve this problem in Sec.~\ref{sec:tableau}.
We present our algorithm in Sec.~\ref{sec:algo_and_analysis} and
illustrate how it may be
used to accommodate quantum hardware with limited connectivity.
Furthermore, we provide estimates of worst-case circuit costs in Sec.~\ref{sec:complexity}.
Additionally, we benchmark the resource costs of our algorithm in Sec.~\ref{sec:results} by considering randomized sets of Pauli operators as well as molecular Hamiltonians, again taking limited qubit connectivity into account.
Finally, we summarize our findings and discuss future avenues for exploring circuit costs with our algorithm in Sec.~\ref{sec:conclude}.

\section{Review of Tableau Representation}
\label{sec:tableau}

Before we propose our algorithm, let us review language frequently used to pose the problem of efficient diagonalization of Pauli operators.
Specifically, we briefly introduce here the tableau representation of Pauli operators, in particular, using the notation in Refs.~\cite{vandenBerg2020circuitoptimization,Murairi:2022zdg}.
We discuss the unitary operations on these Pauli operators in Sec.~\ref{sec:clifford} and pose the problem of diagonalization in Sec.~\ref{sec:problem}.

Suppose we have $N$ commuting Pauli operators acting on $n$ qubits, e.g.,
generating an evolution operator, $U=\exp(\mathrm{i}\sum_{i=1}^Nc_iP_i)$,
or physical observables that we would like to measure, $\{O=\sum_{i=1}^Nc_iP_i\}$.
The tableau corresponding to this collection of Pauli operators $\mathcal{P}=\{P_1,\ldots,P_N\}$ is composed of three arrays: $\mathcal{X}$, $\mathcal{Z}$, and $\mathcal{S}$. The arrays $\mathcal{X}$ and $\mathcal{Z}$ each have dimensions $N \times n$, while the $\mathcal{S}$ array is a column vector with $N$ rows.
For a given Pauli operator $P_i$ with relative phase factor $c_i/|c_i| = \pm 1$, the entries of the three arrays are given by
\begin{equation}
    \begin{split}
    \mathcal{X}\left[i,j\right] &= \begin{cases}
    0 \text{ if the } j\text{th digit of } P_i \text{ is } Z \text{ or } I\\
    1 \text{ otherwise}
    \end{cases},\\
    \mathcal{Z}\left[i,j\right] &= \begin{cases}
    0 \text{ if the } j\text{th digit of } P_i \text{ is } X \text{ or } I\\
    1 \text{ otherwise}
    \end{cases},\\
    \mathcal{S}\left[i\right] &= \begin{cases}
    0 \text{ if } c_i > 0\\
    1 \text{ otherwise}
    \end{cases}.
    \end{split}
\end{equation}

For the remainder of the article, we call the ``tableau'' of our Pauli operators the $N\times 2n$ block matrix $(\mathcal{X}|\mathcal{Z})$.
Here, we neglect the relative phases tracked by the vector $\mathcal{S}$, as they are not necessary for developing a scheme to diagonalize the collection of commuting Pauli operators.
Let us define the following notation for elements of these arrays:
$\mathcal{X}_i$ and $\mathcal{Z}_i$ [with ``lowered'' indices] denote the $i$th column vectors of $\mathcal{X}$ and $\mathcal{Z}$, respectively.
In kind, ``raised'' indices $\mathcal{X}^i$ and $\mathcal{Z}^i$ will denote the $i$th row vectors.
Moreover, a row of the tableau $u=(\mathcal{X}^j,\mathcal{Z}^j)\in\mathbb{F}_2^{2n}$ is said to ``encode'' a Pauli operator $P$ if $P = \bigotimes_{i=1}^n X^{u_i}Z^{u_{n+i}}$ up to an overall phase.

\subsection{Clifford Gates and the Symplectic Group}
\label{sec:clifford}
Since a diagonal Pauli operator may contain only factors $I$ and $Z$, the $\mathcal{X}$ array of diagonal Pauli operators is zero. Consequently, diagonalization of Pauli operators may be viewed as conjugation with unitary operators to reduce the $\mathcal{X}$ array to zero. Specifically, since we seek to transform Pauli operators into other Pauli operators, these unitary operators are Clifford gates, generated by the Hadamard ({\small H}), phase ({\small S}), and controlled-NOT ({\small CNOT}) gates~\cite{Gottesman:1998hu}.

Now, let $C$ be an arbitrary Clifford gate.
Conjugating the Pauli operators of $\mathcal{P}$ with $C$ can equivalently be viewed as an action transforming each column $\mathcal{X}_i$ and $\mathcal{Z}_i$. We denote this action by functions $f_i$ and $g_i$, written
\begin{align}
    \mathcal{X}_i &\mapsto \mathcal{X}^\prime_i = f_i\left(\mathcal{X}_1\,{...}\,\mathcal{X}_N, \mathcal{Z}_1\,{...}\, \mathcal{Z}_N\right)\\
    \mathcal{Z}_i &\mapsto \mathcal{Z}^\prime_i = g_i\left(\mathcal{X}_1\,{...}\,\mathcal{X}_N, \mathcal{Z}_1\,{...}\, \mathcal{Z}_N\right)
\end{align}
which we will show are necessarily linear:
\begin{align}
    f_i(\mathcal{X},\mathcal{Z}) = \sum_{j=1}^n a^{j}_i \mathcal{X}_j + b^j_i \mathcal{Z}_j, \\
    g_i(\mathcal{X},\mathcal{Z}) = \sum_{j=1}^n \Tilde{a}^{j}_i \mathcal{X}_j + \Tilde{b}^{j}_i \mathcal{Z}_j,
\end{align}
where $a^{j}_i$, $b^{j}_i$, $\Tilde{a}^{j}_i$, $\Tilde{b}^{j}_i$ $\in \left\{0,1\right\}$, or, more compactly,
\begin{align}
    f(\mathcal{X},\mathcal{Z}) = \mathcal{X}A + \mathcal{Z}B, \\
    g(\mathcal{X},\mathcal{Z}) =  \mathcal{X}\Tilde{A} + \mathcal{Z}\Tilde{B},
\end{align}
for $A$, $B$, $\Tilde{A}$, $\Tilde{B}$ $\in \mathbb{F}^{n\times n}_2$.
We give the matrices corresponding to elementary Clifford gate conjugations here,
derived from rules presented in, e.g., Refs.~\cite{PhysRevA.70.052328,vandenBerg2020circuitoptimization}. First let $\mathbf{e}_i$ be the unit column vector with $i$th entry equal to 1. Then, we may summarize
\begin{enumerate}[label=(\arabic*)]
\item For a Hadamard gate conjugation on qubit $i$, $\mathrm{H}(i)$, we have corresponding matrices $B=\Tilde{A}=\mathbf{e}_i\mathbf{e}_i^\top$ and $A=\Tilde{B}=\openone_{n\times n}-B$.
\item For a phase gate conjugation on qubit $i$, $\mathrm{S}(i)$, we have corresponding matrices
$A=\tilde{B}=\openone_{n\times n}$, $B=\mathbb{0}_{n\times n}$, and $\tilde{A}=\mathbf{e}_{i}\mathbf{e}_{i}^\top$.
\item For a conjugation with $\text{CNOT}(i,j)$ gates controlled by qubit $i$ and targeting qubit $j$, we have corresponding matrices
$A = \tilde{B}^\top = \openone_{n\times n}+\mathbf{e}_i\mathbf{e}_j^\top$ and $B=\tilde{A}=\mathbb{0}_{n\times n}$.
\end{enumerate}
Since these gates generate the Clifford group, and all compositions of linear functions are also linear, we conclude that all Clifford elements are linear, i.e., are given by some $(A,B,\Tilde{A},\Tilde{B})$.
For example, conjugation with {\small S} on qubits 1 and 2 corresponds to $(\openone, \mathbb{0}, \mathbf{e}_1\mathbf{e}_1^\top+\mathbf{e}_2\mathbf{e}_2^\top, \openone)$, and conjugation with {\small H} on qubit $i$ followed by {CNOT$(i,j)$ corresponds to $(\openone-\mathbf{e}_i\mathbf{e}_i^\top, \mathbf{e}_i\mathbf{e}_i^\top+\mathbf{e}_i\mathbf{e}_j^\top, \mathbf{e}_i\mathbf{e}_i^\top+\mathbf{e}_i\mathbf{e}_j^\top, \openone-\mathbf{e}_i\mathbf{e}_i^\top)$.

Furthermore, we note that the overall matrices
\begin{equation}
    C =
    \begin{pmatrix}
        A & \tilde{A} \\
        B & \tilde{B}
    \end{pmatrix}
\end{equation}
for operations $\mathrm{H}(i)$, $\mathrm{S}(i)$, and $\text{CNOT}(i,j)$ each satisfy the identity
\begin{align}
    C\Lambda C^\top = \Lambda =\begin{pmatrix}
        \mathbb{0}_{n\times n} & \openone_{n\times n}\\
        \openone_{n\times n} & \mathbb{0}_{n\times n}
    \end{pmatrix}.
    \label{eq:eq:symplectic_form}
\end{align}
Therefore, the same identity must immediately hold for all compositions of these gates.
Thus, we explicitly see the connection of Clifford gates to the group of $2n\times2n$ symplectic matrices on $\mathbb{F}_2$, $\mathrm{Sp}(2n,\mathbb{F}_2)$, as noted in Refs.~\cite{Rains:1997uh,2007RSPSA.463.2887K,Koenig:2014wvh}.

\subsection{Statement of the problem}
\label{sec:problem}

Let $n$ be the total number of qubits. Suppose we are given a set of $N$ commuting Pauli operators $\mathcal{P} = \left\{ P_1, \ldots, P_N \right\}$.
We would like to produce a unitary circuit with which we will conjugate the Pauli operators in this set in order to simultaneously diagonalize them.
In the interest of simplifying applications to quantum hardware, we would like to reduce {the overall circuit depth} and the total number of two-qubit gates involved as well as enhance flexibility of diagonalization strategies to accommodate limited qubit connectivity.

It can be shown that such a set may be generated by at most $r\leq\min(n,N)$ Pauli operators.
Let us call $T = \left\{t_1, {...}, t_r\right\}$ a generating set of $\mathcal{P}$ if each element of $\mathcal{P}$ is a product of elements of $T$.
That is, for each $P\in \mathcal{P}$, $P = \prod_{i=1}^{r} t^{\alpha_{i}}_{i}$ for some $\alpha_i \in \left\{0,1\right\}$ up to an overall phase.
In addition, if no element of $T$ can be expressed as a product of other elements of $T$, we say that the elements of $T$ are independent.

Now, let $T$ be an independent generating set of $\mathcal{P}$.
$T$ can be obtained by first constructing the tableau of $\mathcal{P}$ and row reducing the $\mathcal{X}$ and $\mathcal{Z}$ arrays.
Importantly, a basis that diagonalizes $T$ will also diagonalize $\mathcal{P}$.
Meanwhile, since $T$ is an independent set, it is more constrained than $\mathcal{P}$ is.
Therefore, we may reduce our problem to finding a hardware-efficient diagonalization of $T$.

\section{Qubitwise Diagonalization Algorithm}
\label{sec:algo_and_analysis}

In this section, we propose an algorithm for simultaneously diagonalizing Pauli operators generating a desired set.
First, we outline the philosophy of the algorithm in Sec.~\ref{sec:algo_exp}.
More technically, we outline precisely the recursive steps of the proposed algorithm in Sec.~\ref{sec:algorithm}.
In kind, we explain how one can apply these steps in order to accommodate limited qubit connectivity in Sec.~\ref{sec:qconnect}.

\subsection{Approach}
\label{sec:algo_exp}
Our approach to diagonalize $T$ is to iteratively diagonalize all Pauli operators one qubit at a time. In terms of the tableau for $T$, $(\mathcal{X}|\mathcal{Z})$, for some qubit $i$ we solve the equation $\mathcal{X}^\prime_i = \sum_{j=1}^n a_i^j \mathcal{X}_j + b_i^j \mathcal{Z}_j = 0$ for factors $a_i^j$ and $b_i^j$, which we will show prescribes a sequence of Clifford gates with which we conjugate the elements of $T$, rendering all operators diagonal on $i$. In matrix form,
\begin{equation}
    \label{eq:nspace-equation}
    \mathcal{X}^\prime_i = \biggl( \mathcal{X} \; \bigg| \; \mathcal{Z} \biggr)\; \begin{pmatrix}
    \boldsymbol{a}_i\\
    \boldsymbol{b}_i
    \end{pmatrix} = 0,
\end{equation}
where $\boldsymbol{a}_i$ denotes the column vector $\left(a_i^1, \ldots, a_i^n\right)^\top$ and similarly for $\boldsymbol{b}_i$.
We may interpret this null vector as the constraint that certain columns in $\mathcal{X}$ or $\mathcal{Z}$ must sum to a zero vector, i.e., $\sum_{j=1}^n \mathcal{X}_j a_i^j + \mathcal{Z}_j b_i^j=0$.

Equation~\eqref{eq:nspace-equation} has a non-zero solution if the null space of $(\mathcal{X} | \mathcal{Z})$ is non-trivial.
Since $T$ is an independent set of commuting Pauli operators, this tableau forms a matrix of rank $r \leq n$.
Also, $(\mathcal{X} | \mathcal{Z})$ has at most $n$ rows but $2n$ columns. Therefore, this matrix must have a null space of dimension $2n-r \geq n$, and so such a vector $(\boldsymbol{a}_i\, , \boldsymbol{b}_i)$ exists.
We can then obtain a solution $(\boldsymbol{a}_i\, , \boldsymbol{b}_i)$ for example by applying Gauss-Jordan elimination on the matrix $(\mathcal{X} | \mathcal{Z})${\ora, as we will discuss more explicitly in Sec.~\ref{sec:complexity_prelim}}.

{In fact, since the operators considered here are independent and commute, we have $\mathcal{X}^j\cdot\mathcal{Z}^k+\mathcal{Z}^j\cdot\mathcal{X}^k=0$, implying that examples of such vectors in the null space include the $r$ rows of $(\mathcal{Z}|\mathcal{X})=(\mathcal{X}|\mathcal{Z})\Lambda$.
More generally, we can see that a vector $u\in\mathbb{F}_2^{2n}$ is an element of the null space if and only if it encodes a Pauli operator that commutes with each Pauli operator encoded by $(\mathcal{Z}|\mathcal{X})$. ({To see this fact, recall that a vector is in the null space when $(\mathcal{X}|\mathcal{Z})u=0$, or equivalently $(\mathcal{Z}|\mathcal{X})\Lambda u=0$. Writing out $u=(\boldsymbol{v},\boldsymbol{w})$, where $\boldsymbol{v},\boldsymbol{w}\in\mathbb{F}_2^{n}$, it follows that $\mathcal{Z}^k\cdot \boldsymbol{w} + \mathcal{X}^k \cdot \boldsymbol{v}=0$, implying $u$ commutes with the Pauli operator encoded by the row $k$ of the tableau $(\mathcal{Z}|\mathcal{X})$, for each $k=1,\ldots,r$.})
{\ora In terms of classical coding theory~\cite{nielsen_chuang_2010}, we may view the tableau as a parity check matrix for a linear code spanned by these null vectors as code words.}

One can {\ora embed} this null vector $(\boldsymbol{a}_i\, , \boldsymbol{b}_i)$ as the $i$th columns of matrices $A$ and $B$ encoding an overall unitary operation that renders all Pauli operators diagonal on qubit $i$.
Importantly, there is still plenty of freedom in the choices of Clifford operations to specify $\tilde{A}$ and $\tilde{B}$ as well as the other columns of $A$ and $B$.
We exploit this freedom to prescribe the following instructions, which generate a particular choice of such matrices sending $\mathcal{X}_i\mapsto0$.
Additionally, \ora{these degrees} of freedom \cyan{as well as the choice of a suitable null vector} can be further exploited to make choices that maintain low circuit depth given a particular quantum device where qubit connectivity is not all-to-all. \ora{For completeness,} \cyan{we will also show how a suitable null vector can be chosen efficiently.}
}

{\ora
In short: an element of the tableau's null space indicates columns of $\mathcal{X}$ and $\mathcal{Z}$ that are dependent and therefore may be added to diagonalize operators along some qubit $i$ at each stage.
As we detail below, these addition operations imply a concrete list of simple one- and two-qubit gates on a diagonalizing circuit.
}

\subsection{The Algorithm}
\label{sec:algorithm}
Now, we are ready to prescribe the instructions to perform iteratively at each stage. Let $\alpha = 1, \ldots, s\leq n$ be the stage where $n^{(\alpha)}$ digits are not yet diagonal for all Pauli operators.
Moreover, $T^{(\alpha)}$ 
is the corresponding generating set of size $r^{(\alpha)}=|T^{(\alpha)}|$. Stage $\alpha=1$ begins with the initial problem for our algorithm, i.e., $T^{(1)}=T$. 
At each stage $\alpha$, we construct the corresponding tableau $(\mathcal{X}^{(\alpha)}|\mathcal{Z}^{(\alpha)})$.
Then, we select a vector $(\boldsymbol{v}^{(\alpha)}, \boldsymbol{w}^{(\alpha)})$ in the null space of $(\mathcal{X}^{(\alpha)}|\mathcal{Z}^{(\alpha)})$.
Choose a qubit $i$ such that $v^{(\alpha)i} = 1$ or $w^{(\alpha)i} = 1$. 
Pauli operators will be diagonalized on this qubit in two steps by the procedure below:

(1) The first step consists of applying single-qubit Clifford gates to update each column $j=1,\ldots,n^{(\alpha)}$: $\mathcal{X^{(\alpha)}}_j\mapsto\mathcal{X^{(\alpha)}}_j'=v^{(\alpha)j}\mathcal{X^{(\alpha)}}_j+w^{(\alpha)j}\mathcal{Z^{(\alpha)}}_j$, according to the rules relating $a$ and $b$ to corresponding single-qubit Clifford gates outlined in Sec.~\ref{sec:clifford}.
\begin{enumerate}[label=(\alph*)]
\item If 
$w^{(\alpha)j}=0$, then we simply apply the identity gate, i.e., we need not do anything for this $j$.
\item If $v^{(\alpha)j}=0$ and $w^{(\alpha)j}=1$, perform a conjugation with the gate $\mathrm{H}(j)$.
\item If $v^{(\alpha)j}=1$ and $w^{(\alpha)j}=1$, perform a conjugation with the gate $\mathrm{S}(j)$ followed by a conjugation with the gate $\mathrm{H}(j)$.
\end{enumerate}
At the end of this step, we have
\begin{equation}
    \mathcal{X^{(\alpha)}}^{\prime}_j = v^{(\alpha)j} \mathcal{X}^{(\alpha)}_j + w^{(\alpha)j} \mathcal{Z}^{(\alpha)}_j
    \label{eq:step1_result}
\end{equation}
with $v^{(\alpha)j}=1$ or $w^{(\alpha)j}=1$.
This step essentially selects 
all of the columns of $\mathcal{X}^{(\alpha)}$ {and $\mathcal{Z}^{(\alpha)}$}
that are involved in reducing the column $\mathcal{X}^{(\alpha)}_i$ to zero and stores them in $\mathcal{X}^{(\alpha)\prime}$.

(2) The final step is then to
add all of these columns into $\mathcal{X}^{(\alpha)}_i$. We carry out this step using conjugations with {\small CNOT} gates. In particular, for each $j=1,\ldots,n^{(\alpha)}$, if $v^{(\alpha)j} = 1$ or $w^{(\alpha)j} = 1$ and $j \neq i$, perform a conjugation with {CNOT}$(j, i)$. At the end of this step, the columns $\mathcal{X^{(\alpha)}}^{\prime}_j$ selected by having coefficients $v^{(\alpha)j} = 1$ or $w^{(\alpha)j} = 1$ are added to the column $\mathcal{X^{(\alpha)}}^\prime_i$. We therefore obtain $$\mathcal{X^{(\alpha)}}^{\prime\prime}_i = \sum_{j=1}^n v^{(\alpha)j} \mathcal{X}^{(\alpha)}_j + w^{(\alpha)j} \mathcal{Z}^{(\alpha)}_j=0,$$ which is guaranteed to equal zero as desired since $(\boldsymbol{v}^{(\alpha)}, \boldsymbol{w}^{(\alpha)})$ is in the null space of $(\mathcal{X}^{(\alpha)}|\mathcal{Z}^{(\alpha)})$; see Eq.~\eqref{eq:nspace-equation} and the discussion below it.

Thus concludes stage $\alpha$. After we have diagonalized the Pauli matrices acting on qubit $i$ with unitary operations, the Pauli operators acting on the remaining qubits must also commute.
Therefore, we can diagonalize the remaining qubits in recursive stages.

After stage $\alpha$ is complete, we have $n^{(\alpha + 1)} \leq n^{(\alpha)} - 1$.
{Note $n^{(\alpha+1)}$ is not necessarily $n^{(\alpha)}-1$, as other qubits beside $i$ may have been diagonalized in the process.
Proceeding to the next stage, to obtain $T^{(\alpha + 1)}$, we exclude all the qubits that are acted upon only by $I$ or $Z$.
Consequently, note that $r^{(\alpha+1)}\leq r^{(\alpha)}$, as the number of distinct generators may decrease after reducing the number of qubits still in consideration.
Then, we repeat the above process for $T^{(\alpha + 1)}$.
At a stage $\alpha=s$ where $n^{(s)} = 1$, $r^{(s)} = 1$ and the step 1 of the procedure above is sufficient to diagonalize the qubit.
Thus, the recursion indeed terminates.
The complete algorithm is summarized below in pseudocode.

\input{diagonalization_pseudocode}

Note that the example of a simple construction for step 2 above provides a circuit with depth linear with the number of {\small CNOT} gates at each stage, assuming hardware on which commuting {\small CNOT} gates that share qubits \emph{cannot} be performed in parallel.
In Section~\ref{sec:depth}, we discuss another construction with circuit depth that grows only logarithmically with the number of {\small CNOT} gates at each stage.
Via this latter construction we
achieve the ${O}(n \log r)$ overall circuit depth advertised. \\

\subsection{Accommodating Qubit Connectivity}
\label{sec:qconnect}
We now turn our attention to discussing how this algorithm may be {\ora tailored to} hardware with limited connectivity, in the sense that we would like it to involve relatively few {\small SWAP} gates.
Suppose that the graph $G = (V,E)$ represents the hardware connectivity. That is, a qubit $q_i$ is represented by a vertex in $V$, and there is an edge $(q_i, q_j) \in E$ if qubits $q_i$ and $q_j$ are directly connected on the hardware (i.e., permitting a {\small CNOT} operation with either qubit as target or control and without {\small SWAP} operations).\footnote{One may also consider hardware with yet further limited connectivity such that a gate CNOT$(i,j)$ may be performed but not CNOT$(j,i)$. The remainder of this discussion below may be generalized to this case, however potentially requiring yet additional resources to accommodate such limitations. }
Qualitatively, the strategy is to exploit the degrees of freedom still allowed by the algorithm described in the previous section, particularly in the choices of the null vector and the order of additions in tableau columns prescribed by this null vector.
We elaborate on how these choices can be related and how they can be made with limited qubit connectivity in mind.

The first task is to choose a null space vector $(\boldsymbol{v}, \boldsymbol{w})$ that is compatible with the qubit connectivity of a given quantum device.
However, we stress that the choice of this vector also entails a list of qubits, for pairs of which ($q_i$ and $q_j$) we add corresponding column vectors of the tableau, via {\small CNOT} gates (and potentially {\small SWAP} gates as well).
We would like to choose qubit pairs whose paths involve relatively few {\small SWAP} gates to be realized.
That is, for all $i$ such that $v^i = 1$ or $w^i = 1$ and $j$ such that $v^j = 1$ or $w^j = 1$, we desire a relatively short path between vertices $q_i$ and $q_j$ in the connectivity graph $G$.
As such, we can choose a null vector for which the qubits $q_i$ and $q_j$ are relatively close to one another in $G$.
Furthermore, we give below an explanation of how, given an arbitrary null space vector, we can choose pairs of qubits on which we will perform {\small CNOT} gates, in order to further accommodate limited connectivity.

Now given an arbitrarily chosen null vector, we can describe the problem of choosing how one adds columns as per step 2 for this null vector.
(We may assume that step 1 of the algorithm is already completed, as it involves only one-qubit gates.)
For step 2, let us denote the qubits where $(\boldsymbol{v}, \boldsymbol{w})$ is nonzero as $Q = \{ q_1, q_2, q_3, \ldots\}$ and
the columns $\mathcal{X}_q$ with $q \in Q$
to be added using {\small CNOT} gate conjugations.
In general, since the pairs $(q_i, q_j)\subseteq V$ may be disconnected on the hardware in consideration, we would need to perform appropriate conjugations with {\small SWAP} gates resulting in $q_i \mapsto q^{\prime}_i$ and $q_j \mapsto q^{\prime}_j$ such that $q^{\prime}_i$ and $q^{\prime}_j$ are connected on the hardware (i.e., share an edge in $E$).
This transformation can be accomplished along any path in $G$ connecting $q_i$ and $q_j$, and the number of {\small SWAP} gates depends on the length of the path.
As such, we seek to minimize the length of a path that traverses all of the qubits $q \in Q$. \cyan{Our minimization problem is similar to the well-known traveling salesperson problem (TSP)~\cite{Karp1972}. However, in contrast with the TSP, which requires every node be visited exactly once, our problem requires only the nodes in $Q$ be visited once. That is, a node not in $Q$ may or may not be visited at all.}
This minimization problem \cyan{naturally} maps to the so-called shortest-path problem with specified nodes (SPPSN)\footnote{\ora{In other words,} \cyan{the problem is to find the shortest path passing through a subset of vertices in a graph.}}~\cite{10.2307/2582232}, \cyan{where the specified nodes are the nodes in $Q$}.
\cyan{This problem has been extensively studied, and several solutions are provided in, e.g., Refs.~\cite{10.2307/2582232,laporte,dreyfus,ibaraki}.}
Indeed, solving the SPPSN gives us a sequence of qubits $\{q_{\sigma1}, q_{\sigma2}, q_{\sigma3}, {...}\}$ along which the number of {\small SWAP} gates is minimized when performing step 2.
Finally, we note that the choice of null vector essentially dictates which SPPSN we must solve.

As explained above, modifications in step 2 of our algorithm can allow for multiple avenues to reduce the {\small SWAP} gates required.
Not only can we draw upon efficient solutions to the related generalization of a TSP (i.e., choose how to add columns of $\mathcal{X}'$), but also we are afforded a choice of which such problem we would like to solve (i.e., choose which columns of $\mathcal{X}'$ to be added according to $(\boldsymbol{v},\boldsymbol{w})$).
In Sec.~\ref{sec:result-connectivity}, we demonstrate how this strategy can be adapted to hardware with linear connectivity.

\section{Analysis of the Algorithm}
\label{sec:complexity}

Here, we estimate the size of quantum circuits produced by the algorithm presented in the previous section.
In particular, we focus on the overall circuit depth as well as the number of {\small CNOT} gates entailed.
First, we explain in Sec.~\ref{sec:complexity_prelim} how these qualities of the circuit are related to the null vectors introduced in Sec.~\ref{sec:algo_exp}.
With this understanding in hand, we then can estimate the circuit depth (Sec.~\ref{sec:depth}) and total two-qubit gate count (Sec.~\ref{sec:count}) required to implement the diagonalization circuits prescribed by our algorithm.

\subsection{Preliminaries}
\label{sec:complexity_prelim}

Before we proceed, {\ora it is helpful to recall the perspective that our algorithm essentially searches} for $\mathcal{X}$ and $\mathcal{Z}$ columns that are dependent and adds them to diagonalize along some qubit $i$ at each stage. This interpretation is useful in estimating the cost.

\begin{definition}
    The symplectic weight of a vector $({v}, {w}) \in \mathbb{F}^{2n}_2$ is defined as
    \begin{equation}
        \omega(\boldsymbol{v},\boldsymbol{w}) = \left|\left\{i\mid(v^i,w^i) \neq (0,0) \right\}\right|.
    \end{equation}
    \label{def:symplectic-weight}
\end{definition}
In other words, the symplectic weight counts the number of digits not equal to $I$ in the corresponding Pauli operator {\ora (i.e., the operator's Hamming weight)}.
We will also find $\omega(\boldsymbol{v},\boldsymbol{w})$ of null space vectors helpful in estimating the number of {\small CNOT} gate conjugations.

\begin{lemma}
    Given a vector $(\boldsymbol{v}, \boldsymbol{w})$ in the null space of $(\mathcal{X}^{(\alpha)}|\mathcal{Z}^{(\alpha)})$, our algorithm prescribes $\omega(\boldsymbol{v},\boldsymbol{w}) - 1$ {\small CNOT} gate conjugations to diagonalize one qubit.
    \label{th:cnot-cost-symplectic-weight}
\end{lemma}
Suppose qubit $i$ is to be diagonalized using the vector $(\boldsymbol{v}, \boldsymbol{w})$ through steps 1 and 2.
By Definition \ref{def:symplectic-weight}, step 1 will update $\omega(\boldsymbol{v}, \boldsymbol{w})$ columns in $\mathcal{X}$.
Then, step 2 will use $\omega(\boldsymbol{v}, \boldsymbol{w}) - 1$ {\small CNOT} conjugations, because there are $\omega(\boldsymbol{v}, \boldsymbol{w}) - 1$ columns in $\mathcal{X}$ to be added to column $\mathcal{X}_i$.

\begin{lemma}
    The number of {\small CNOT} gate conjugations required to diagonalize a given qubit at stage $\alpha$ is at most $r^{(\alpha)}$ if $n^{(\alpha)} > r^{(\alpha)}$, and at most $r^{(\alpha)} - 1$ if $n^{(\alpha)} = r^{(\alpha)}$.
     \label{th:cnot-stage}
\end{lemma}
We first consider the case $n^{(\alpha)} > r^{(\alpha)}$.
At stage $\alpha$, the rank of $(\mathcal{X}^{(\alpha)}|\mathcal{Z}^{(\alpha)})$ is $r^{(\alpha)}$, implying that any collection of $r^{(\alpha)} + 1$ columns is a dependent set.
Therefore, there exists a vector $(\boldsymbol{v}, \boldsymbol{w})$ in the null space of $(\mathcal{X}^{(\alpha)}|\mathcal{Z}^{(\alpha)})$ with weight $\omega(\boldsymbol{v}, \boldsymbol{w}) \leq r^{(\alpha)} + 1$.
Consequently, using Lemma~\ref{th:cnot-cost-symplectic-weight}, we demand at most $r^{(\alpha)}$ {\small CNOT} gate conjugations at this stage.

For the case $n^{(\alpha)} = r^{(\alpha)}$, we know $\omega(\boldsymbol{v}, \boldsymbol{w}) \leq r^{(\alpha)}$ for all $(\boldsymbol{v},\boldsymbol{w})$ in the null space, because there are only $r^{(\alpha)}$ qubits. Therefore, this case entails at most $r^{(\alpha)} - 1$ {\small CNOT} gate conjugations.

Finally, we stress that, in either case, such a null space vector can be produced efficiently.
For example, we can find such a null space vector by finding a set of $\leq r^{(\alpha)}+1$ columns adding to zero.
To demonstrate this {\ora bound}, we first transform the matrix $(\mathcal{X}^{(\alpha)} | \mathcal{Z}^{(\alpha)})$ to its reduced row-echelon form in polynomial complexity~\cite{KOC1991118}.
Then, up to a relabeling of qubits, our tableau can be written in the so-called standard form of a stabilizer code (see e.g.~\cite{nielsen_chuang_2010}):
\begin{equation}
    \begin{pmatrix}
        \openone_{r^{(\alpha)} \times r^{(\alpha)}} & \mathcal{A}_{r^{(\alpha)} \times (n^{(\alpha)} - r^{(\alpha)})} & | & \mathcal{B}_{r^{(\alpha)} \times n^{(\alpha)}}
    \end{pmatrix}
\end{equation}
where
$\mathcal{A}$ and $\mathcal{B}$ are matrices over $\mathbb{F}_2$ with the indicated dimensions.
{\pp
In this form, any column of $\mathcal{A}$ or $\mathcal{B}$ is clearly a linear combination of the columns of $\openone$;
the $m$th column $c_m$ ($r^{(\alpha)} <m\leq 2n$) of the transformed tableau above can be written as
$c_m=\sum_{i=1}^{r^{(\alpha)}}c^i_m\mathbf{e}_i$.
In other words, any column of $\mathcal{A}$ or $\mathcal{B}$ may be added with a subset of the columns of $\openone$ to yield a zero vector, and so a vector encoding this sum would be a suitable null vector to start our algorithm of Sec.~\ref{sec:algorithm}.
Explicitly, the collection of these columns to be summed corresponds to a null vector $u\in\mathbb{F}_2^{2n^{(\alpha)}}$ with $u^i=1$ and $u^m=1$ only at $i$ and $m$ for which $c^i_m=1$.
In fact, one may choose the column $c$ of $\mathcal{A}$ or $\mathcal{B}$ with the lowest (Hamming) weight to reduce the weight of the corresponding null vector $\omega(u)$ as well as the gate cost that it entails (as per Lemma~\ref{th:cnot-cost-symplectic-weight}).
}

 \subsection{Circuit Depth}
 \label{sec:depth}
Now, we will analyze the overall circuit depth of the algorithm.
Notably, on hardware where commuting {\small CNOT} gates can be performed in parallel, as discussed in e.g., Ref.~\cite{v001a005}, we in fact already have constant depth at each stage via the instructions provided in Sec.~\ref{sec:algorithm}, yielding overall depth ${O}(n)$.
Remarkably, accommodating more generic hardware, we find that the circuit has depth ${O}(n\log{r})$ in contrast with the number of gates, which we will show is ${O}(nr)$.

At each stage $\alpha$, we note that step 1 produces a circuit of depth at most two, because only single-qubit gates may be performed on each qubit.
Meanwhile, as we shall argue below, the {\small CNOT} gates conjugations in step 2 can be performed with depth ${O}(\log r^{(\alpha)})$.
Moreover, since there are at most $n$ recursive stages in the algorithm, the overall circuit has depth ${O}(n \log r)$.

Recall that step 1 modifies the columns involved in the diagonalization according to Eq.~\eqref{eq:step1_result}, and step 2 adds these $\mathcal{X}^\prime$ columns.
Without loss of generality, let us denote these columns by $\mathcal{X}^\prime_1, \mathcal{X}^\prime_2, {...}, \mathcal{X}^\prime_{\omega}$ where $\omega$ is the weight of the null space vector used for diagonalization.
In step 2, we can add these columns in pairs sequentially (as described earlier in Sec.~\ref{sec:algorithm}), or alternatively we may form disjoint pairs to be added in parallel to the same end.
That is, we can perform {CNOT$(2j, 2j-1)$} resulting in $\mathcal{X}^{\prime}_{2j-1} \leftarrow \mathcal{X}^\prime_{2j-1} + \mathcal{X}^\prime_{2j}$ for $j \in \{1, 2, \ldots, \lfloor \omega/2\rfloor\}$.
Note that all of these {\small CNOT} gates amount to depth 1, since they each act on totally different qubits.
Now, we add these resulting columns $\mathcal{X}^{\prime}_1, \mathcal{X}^{\prime}_3, \ldots$ in the same fashion.
This process repeats until we are left with only one $\mathcal{X}$ column, and each round of this pairwise addition process is a circuit with depth 1.
We summarize this procedure in pseudocode in Algorithm~\ref{alg:diag2}, replacing lines 21--25 of the instructions given in Algorithm~\ref{alg:diag}.

\input{modified_step2_pseudocode}

\noindent Since each round of pairwise additions reduces the number of $\mathcal{X}$ columns to be added by about one half, there are ${O}(\log_2 \omega)$ such rounds, each with depth 1.
Using Lemma~\ref{th:cnot-stage}, we can choose $\omega  \leq r^{(\alpha)} + 1$, and so each stage $\alpha$ involves a circuit with depth ${O}(\log r^{(\alpha)})$.

 \subsection{{\small CNOT} Gate Count}
\label{sec:count}

 Having analyzed the circuit depth, we now turn our attention to discussing the total number of {\small CNOT} gates.

 \begin{theorem}
     The number of {\small CNOT} gate conjugations required by the algorithm (Sec.~\ref{sec:algorithm}) is at most $n\,r - \frac{r(r+1)}{2}$.
     \label{th:weak-bound}
 \end{theorem}
To establish this bound, we will use Lemma \ref{th:cnot-stage} to count the number of {\small CNOT} gate conjugations at each stage.
The first stage, $\alpha = 1$, requires at most $r^{(1)} = r $ {\small CNOT} gate conjugations.
For the subsequent stage, $\alpha = 2$, if $n^{(2)} \geq r$, the number of independent generators at $r^{(2)}$ may remain $r$.
Consequently, for all stages $\alpha$ such that $n^{(\alpha)} > r$, the inequality $r^{(\alpha)} \leq r$ holds.
There are at most $\left(n - r \right)$ such stages, amounting to at most $\left(n - r \right) r$ {\small CNOT} gate conjugations.
Once we reach a stage $s$ such that $n^{(s)} = r$, the stage $\alpha = s + j$ has $r^{(s + j)} \leq n^{(s+j)} \leq r - j$, because each stage diagonalizes on at least one qubit.
The number of {\small CNOT} gate conjugations in these stages is therefore at most $\sum_{j=0}^{r-1} (r - j - 1) = r(r - 1)/2$.
Putting both {\small CNOT} gate conjugation counts together, we obtain $n\,r - r(r+1)/2$.

This bound is also derived for a different algorithm in Ref.~\cite{Crawford2021efficientquantum}. Notably, in the limiting case $r = n$, this bound reduces to the known result $n(n-1)/2$ (see, e.g., Refs.~\cite{vandenBerg2020circuitoptimization,Miller:2022sol}).

It is important to note that our algorithm guarantees this bound without further classical optimization.
That is, one may select any $r^{(\alpha)} + 1$ columns to form a dependent set at each stage $\alpha$, and this upper bound will hold.
Now, we demonstrate that we can achieve a lower estimate of {\small CNOT} cost by minimizing the size of the dependent set of columns we select in each stage, entailing a higher classical complexity cost.

\begin{lemma}
     A qubit can be diagonalized at stage $\alpha$ with at most $\lfloor \frac{r^{(\alpha)}}{2} \rfloor$ {\small CNOT} gate conjugations.
     \label{th:tight-bound-stage-alpha}
 \end{lemma}
In exchange for a lower gate cost, one may accept a large classical complexity by searching the entire null space for the vector with the lowest symplectic weight by Lemma~\ref{th:cnot-cost-symplectic-weight}. Below, we will use the properties of the null space of $(\mathcal{X}| \mathcal{Z})$ at each stage to establish a stricter lower bound on the number of {\small CNOT} gates.

The rows of $(\mathcal{Z}^{(\alpha)}|\mathcal{X}^{(\alpha)})$ span a $r^{(\alpha)}$-dimensional subspace of the null space, due to the fact that the Pauli operators represented by each row of $(\mathcal{X}^{(\alpha)}|\mathcal{Z^{(\alpha)}})$ commute with one another. 
Moreover, the Pauli operators in $(\mathcal{Z}^{(\alpha)}|\mathcal{X}^{(\alpha)})$ may be viewed as a $[[n^{(\alpha)},k,d]]$ stabilizer code. The code maps $k=n^{(\alpha)}-r^{(\alpha)}$ {\ora logical} qubits to $n^{\ora(\alpha)}$ physical qubits with the capability of correcting at most $(d-1)/2$ errors.
Moreover, we call $S$ the stabilizer generated by the Pauli operators represented by the rows of $(\mathcal{Z}^{(\alpha)}|\mathcal{X}^{(\alpha)})$. For a more general review, see, e.g., Refs.~\cite{1997PhDT.......232G,nielsen_chuang_2010}.

Let $C(S)$ be the centralizer of $S$, the Pauli operators that commute with each element of $S$. For stabilizer codes, it is known that the normalizer is identical to the centralizer $N(S) = C(S)$ \cite{1997PhDT.......232G}. In Sec.~\ref{sec:algo_exp}, we argued that a vector $u$ commutes with every element of $(\mathcal{Z}^{(\alpha)}|\mathcal{X}^{(\alpha)})$, i.e., is an element of $N(S)$, if and only if it is in the null space of $(\mathcal{X}^{(\alpha)}|\mathcal{Z}^{(\alpha)})$.
Therefore, the task of finding a null space vector $u$ of weight $\omega$ is equivalent to searching $N(S)$ for a Pauli operator of the desired weight. On the other hand, the quantum Singleton bound~\cite{PhysRevA.55.900,Rains:1997uh,Cerf:1997vf} directly implies that there exists a Pauli operator in $N(S)$ with weight $\omega = d \leq \lfloor r^{(\alpha)}/2 \rfloor +1$. Finally, invoking Lemma~\ref{th:cnot-cost-symplectic-weight} concludes the proof.

\begin{theorem}
     The number of {\small CNOT} gate conjugations required by the algorithm (Sec.~\ref{sec:algorithm}) is at most $n\lfloor \frac{r}{2}\rfloor - \lfloor\frac{r}{2}\rfloor^2$.
     \label{th:tighter-bound}
 \end{theorem}
To achieve this bound, at each stage $\alpha$, we select the vector with weight $d\leq \lfloor r^{(\alpha)}/2 \rfloor + 1$ promised by Lemma~\ref{th:tight-bound-stage-alpha}. Then, we follow an argument similar to that of Theorem~\ref{th:weak-bound} to find the total cost.
That is, while $n^{(\alpha)} \geq r$, we use at most $\lfloor r/2 \rfloor$ {\ora {\small CNOT} gate conjugations}. Since there are at most $n - r + 1$ of such stages, they amount to at most $(n - r + 1)\lfloor r/2 \rfloor $ {\small CNOT} conjugations.
After the stage $s$ where $n^{(s)} = r$, we use at most $\lfloor (r-i)/2 \rfloor$ for each subsequent stage $s + i$. 
These stages amount to at most $\sum_{i=1}^{r-1} \lfloor (r-i)/2 \rfloor = \lfloor r/2 \rfloor \lfloor (r-1)/2 \rfloor$ {\small CNOT} conjugations.
Putting both contributions together, we obtain at most
$(n - r + 1)\lfloor r/2 \rfloor + \lfloor r/2\rfloor \lfloor (r-1)/2 \rfloor$
{\small CNOT} conjugations, which
gives us the bound claimed.
\\

{\ora
In the above argument, we have shown an equivalence between finding the distance of a quantum stabilizer code and reducing the {\small CNOT} count at one stage of our qubitwise diagonalization to $\sim r^{(\alpha)}$.
Finding the distance of a code is known to be NP-hard~\cite{641542}, and so we may expect to practically find the operator(s) of the normalizer $N(S)$ with the lowest weight only for a relatively small number of qubits ($n\lesssim30$ on ordinary computers).
}

{\ora
Equivalently, this problem may be restated as minimizing
\begin{align}
    \min_{\{b_1+\cdots+b_{2n-r}\leq z\}}
    \omega\left( \sum_{i=1}^{2n-r} b_i N^i \right),
\end{align}
where $N^i$ are vectors encoding the independent generators of $N(S)$, each $b_i=0$ or $1$, and the sum between $b_iN^i$ vectors is Boolean (i.e., {\small XOR}) and the sum over $b_i$ is decimal.
Here, we must take $z=2n-r$ to search the entirety of $N(S)$ for the lowest-weight solution. 
Nevertheless, a search through linear combinations of fixed sizes $\leq z$ is still polynomial in $n$ and still produces a vector of weight at least as low as that obtained from choosing among $N^i$.
}

{\ora
On the other hand, it is crucial to note that there is a `greediness' to optimizing the (symplectic) weight of our null vector at each stage of our algorithm, as opposed to optimizing the choice for a circuit depth overall.
In the sense, completely optimizing at this stage might not always produce the overall lowest circuit depth after diagonalizing over \emph{all} qubits of our operators.
In the following section, we will discuss further comparisons between circuit depths obtained with null vectors chosen via Gaussian elimination sans optimization (see Lemma~\ref{th:cnot-stage} and its discussion below) and those chosen via complete optimization at each stage.
}

\section{Examples}
\label{sec:results}

In this section, we evaluate the performance of our algorithm for a variety of example applications.
In particular, we use a proposed method of randomized benchmarking
\cite{vandenBerg2020circuitoptimization}
to assess the circuit cost resulting from our algorithm under various circumstances in Sec.~\ref{sec:randbench}.
Secondly, in Sec.~\ref{sec:molecules}, we assess the algorithm's performance in diagonalizing molecular Hamiltonians.
Finally, we provide an example molecular Hamiltonian for which our algorithm accommodates limited qubit connectivity relatively well in Sec.~\ref{sec:result-connectivity}.

\subsection{Random Sets of Commuting Pauli Operators}
\label{sec:randbench}

\begin{figure}[tbp]
  \centering
  \includegraphics{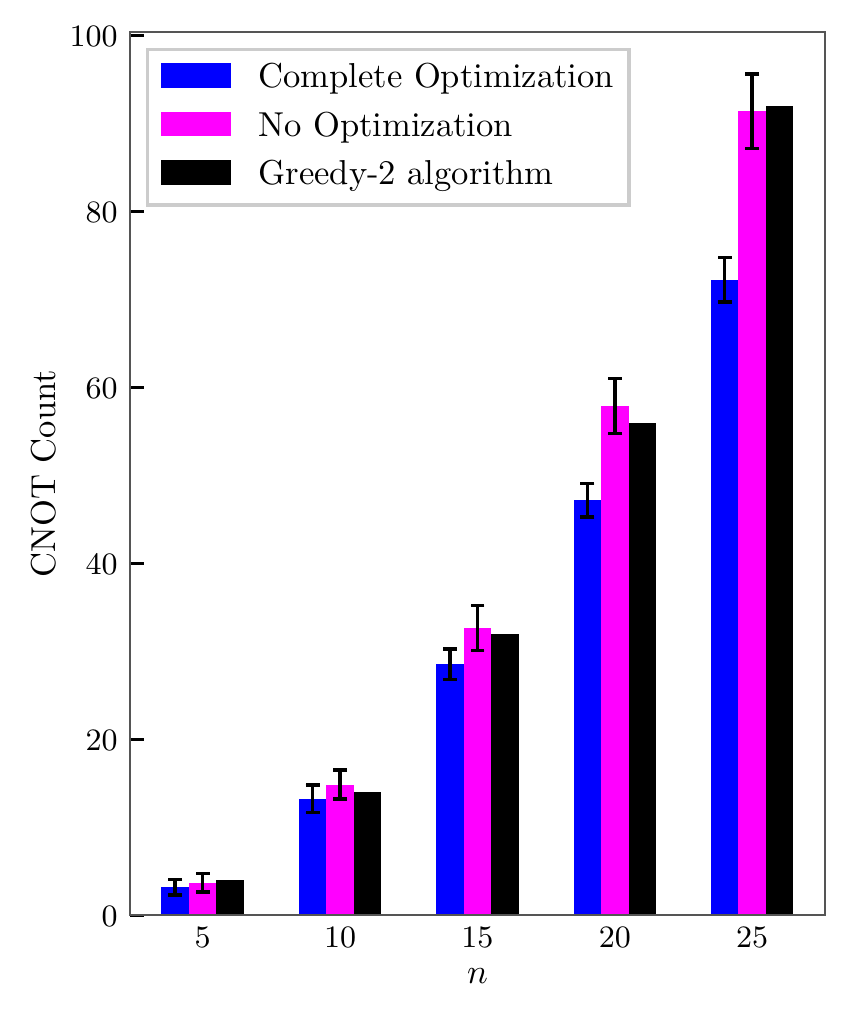}
  \caption{The number of {\footnotesize CNOT} conjugations involved in diagonalizing $n$ commuting Pauli operators acting on $n$ qubits via our qubitwise diagonalization strategy, averaged over $100$ sets generated randomly.
  Each uncertainty bar is the square root of the sample variance of the {\ora$100$} sets.
{\ora We compare gate counts found with and without minimizing the circuit depth, iteratively at each stage of our algorithm (see discussion following Theorem~\ref{th:tighter-bound}). }
  In addition, the magenta (right) bars are in each case the smallest gate counts obtained in Ref.~\cite{vandenBerg2020circuitoptimization}, via the ``greedy-2'' algorithm.
  }
  \label{fig:us_vs_ewout}
\end{figure}

In this subsection, we apply our diagonalization algorithm to sets of commuting Pauli operators generated randomly.
We use the algorithm constructed in Ref.~\cite{vandenBerg2020circuitoptimization} to generate a sample of $100$ distinct sets, each containing $n$ Pauli operators acting on $n$ qubits. In our implementation, we use both the classically efficient algorithm outlined below Lemma~\ref{th:cnot-stage} (``no optimization'') to select a null space vector of weight $\leq r$ and an exhaustive search through the null space to select the vector of lowest weight (``complete optimization'').
For each value of $n$, we report the average number of {\small CNOT} conjugations needed over the $100$ sets.
The error bars in our results are given by the square root of the sample variance. We also superimpose our results to the lowest results obtained in Ref.~\cite{vandenBerg2020circuitoptimization} in Fig.~\ref{fig:us_vs_ewout}.

{\ora
In particular, we can see that there is a marked reduction in the quantum circuit cost when searching for a low-weight null vector to prescribe each stage of our algorithm, at least for totally randomized sets of Pauli operators.
In other words, here we see a trade off, between classical complexity of producing a low-weight null vector and resulting the quantum circuit complexity.} \cyan{Particularly, for $r \ll n$, this classical complexity may become highly prohibitive since the null space is of dimension $2n-r$, in which case the no optimization case conveniently produces a vector of (low) weight at most $r$ anyhow.}
{\ora
For the rest of this section, to assess the costs on a quantum computer of our algorithm while keeping classical complexity low, we limit our algorithm to no optimization (i.e., selecting null vectors via the procedure outlined under Lemma~\ref{th:cnot-stage}).
}

\begin{figure}[t]
    \centering
    \includegraphics{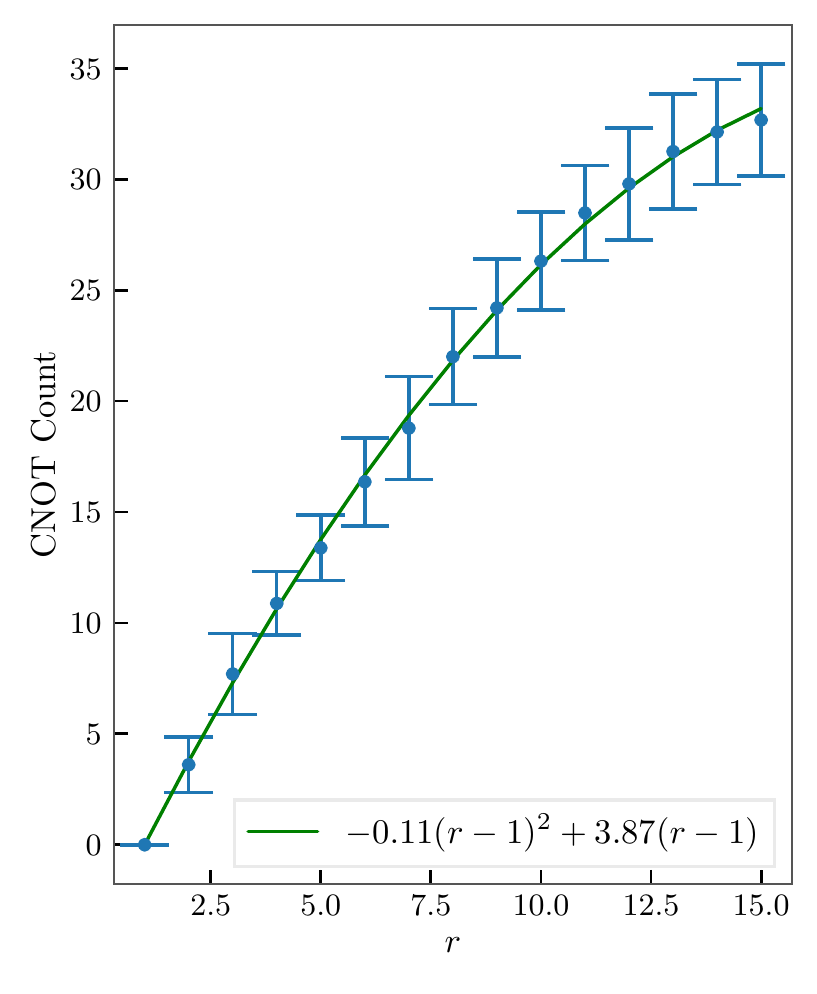}
    \caption{The number of {\footnotesize CNOT} conjugations obtained in diagonalizing a set of $r$ independent commuting Pauli operators acting on {\ora$n=15$} qubits \cyan{using the procedure outlined below Lemma~\ref{th:cnot-stage} (no optimization)}. The data points are averages over {\ora100} randomly generated sets. The uncertainty bars displayed are the square roots of the variances.
    }
    \label{fig:fixed_n}
\end{figure}

Now, we illustrate how our number of {\small CNOT} gates also depends on $r$, the number of independent generators of a set of commuting Paulis.
We first fix $n$, the number of qubits, and again
consider randomly generated sets of $r$ independent commuting Pauli operators.
To obtain a set with only $r \leq n$ operators, we randomly select $n - r$ operators to delete from the set.
Then, we construct diagonalization quantum circuits and report the resulting number of {\small CNOT} gates. Figure~\ref{fig:fixed_n} shows the results for $n = 10$.
As expected per our discussion in Sec.~\ref{sec:count}, we find that, for fixed $n$, gate count follows a quadratic trend in $r$ as $f(r)=-ar^2+br$ [up to a constant shift] for $a,b>0$.

\subsection{Molecular Hamiltonians}
\label{sec:molecules}

\begin{table*}[htbp]
\caption{Results of {\ora qubitwise diagonalization of} molecular Hamiltonians. Throughout the diagonalization, the null space vector at each stage is chosen using the (classically efficient) algorithm described below Lemma~\ref{th:cnot-stage}.
In each row, $n$ and $N$ are the number of qubits and the total number of Pauli operators in each Hamiltonian, respectively. $\kappa$ is the number of sets of commuting Pauli operators, while $r$ is the average number of independent generators per commuting set.
The columns {\footnotesize CNOT} and Depth list the average numbers of {\footnotesize CNOT} gate conjugations and the average quantum circuit depths obtained, respectively, with averages taken over the different sets of commuting Pauli operators in the Hamiltonian.
Finally {\footnotesize CNOT} SD and {Depth SD} are the standard deviations of the corresponding columns.}
\label{tab:molecules}
\begin{tabular}{lrrrrrrrr}
\hline
\hline
Molecule & $n$ & $N$ & $\kappa$ & $r$ & {\footnotesize CNOT} & {\footnotesize CNOT} SD & Depth & Depth SD \\
\hline
HeH$^+$ & 4 & 26 & 3 & 4.00 & 2.00 & 1.63 & 3.33 & 3.40 \\
LiH & 12 & 630 & 25 & 9.00 & 7.44 & 3.42 & 11.12 & 4.34 \\
BeH$_2$ & 14 & 665 & 19 & 11.16 & 9.00 & 3.28 & 11.68 & 4.22 \\
NH$_3$ & 16 & 2296 & 80 & 13.10 & 21.44 & 8.34 & 22.49 & 7.44 \\
BH$_3$ & 16 & 2496 & 97 & 12.01 & 18.21 & 6.36 & 21.26 & 6.64 \\
\hline
\hline
\end{tabular}
\end{table*}

Having assessed the algorithm's performance in diagonalizing commuting Hamiltonians generated randomly, we will now consider its performance for a collection of molecular Hamiltonians as more practical examples.
In particular, we consider one ionic compound HeH$^{+}$ as well as the molecules LiH, BeH$_{2}$, NH$_{3}$, and BH$_3$.

The Hamiltonians
as linear combinations of Pauli operators
are obtained from Pennylane~\cite{2018arXiv181104968B} Quantum Chemistry datasets.
For each of these species, we use the STO-3G basis, and the optimal bond length given in the dataset. Given a Hamiltonian, before diagonalization we must divide the Pauli operators into sets of mutually commuting operators.
As in
Refs.~\cite{Murairi:2022zdg,vandenBerg2020circuitoptimization,Tomesh:2021pns}, this problem can be reduced to a graph-coloring problem, and we use the independent set strategy in NetworkX~\cite{networkx_2008} to color the graph.

Table~\ref{tab:molecules} shows the results of diagonalization of each of these Hamiltonians.
In each case, we observe that both the average number of {\small CNOT} gate conjugations and the circuit depth tend to be near once or twice the number of independent operators generating each set ($r \leq n$).
However, it is worth noting that different choices of strategies to form commuting sets of Pauli operators (while using the same diagonalization algorithm) may allow for higher or lower average gate counts or depths per set.

{\ora
Empirically we find that, for these more physically typical Hamiltonian models rather than randomized sets, in fact a lower circuit complexity is obtained from no optimization rather than complete optimization.
From these cases, we may reasonably suspect that, in general, relatively local and symmetrical Hamiltonians modeling physical systems are more amenable to this simple classical procedure for producing circuit instructions than 
arbitrarily non-local and asymmetrical Hamiltonians of randomized benchmarking.
}

\subsection{Linear Qubit Connectivity}
\label{sec:result-connectivity}

Further, we demonstrate the strategy presented in Sec.~\ref{sec:qconnect} assuming a hardware with linear connectivity. As a proof of principle, we use the ${\rm H}_4$ chain Hamiltonian derived in Ref.~\cite{Miller:2022sol}. This Hamiltonian was obtained by setting the inter-atomic distance $\Delta = 1.0$ \AA\, and using the STO-3G basis.
The Hamiltonian is then mapped to eight-qubit Pauli operators using the Bravyi-Kitaev transformation~\cite{BRAVYI2002210}.
To compare, we use the same division of the Hamiltonian into ten collections of commuting Pauli operators.

First, we consider the set of {\ora 20} Pauli operators shown in Fig.~15 of Ref.~\cite{Miller:2022sol}. There, the authors derived a circuit with 13 {\small H} gates, 18 {\small CNOT} gates, 21 {\small CZ} gates, and 66 {\small SWAP gates} assuming a hardware with linear connectivity.
Via our algorithm of Sec.~\ref{sec:algo_and_analysis}, we can produce a
simpler circuit, with only seven {\small H} gates, five {\small CNOT} gates, and 11 {\small SWAP} gates. 
Figures~\ref{fig:qfull} and~\ref{fig:qlinear} show the diagonalization circuit assuming fully connected hardware and linearly connected hardware, respectively.

\begin{figure*}[htbp]
    \centering
     \subfloat[\label{fig:qfull}]{
     \includegraphics[width=0.4\textwidth]{
     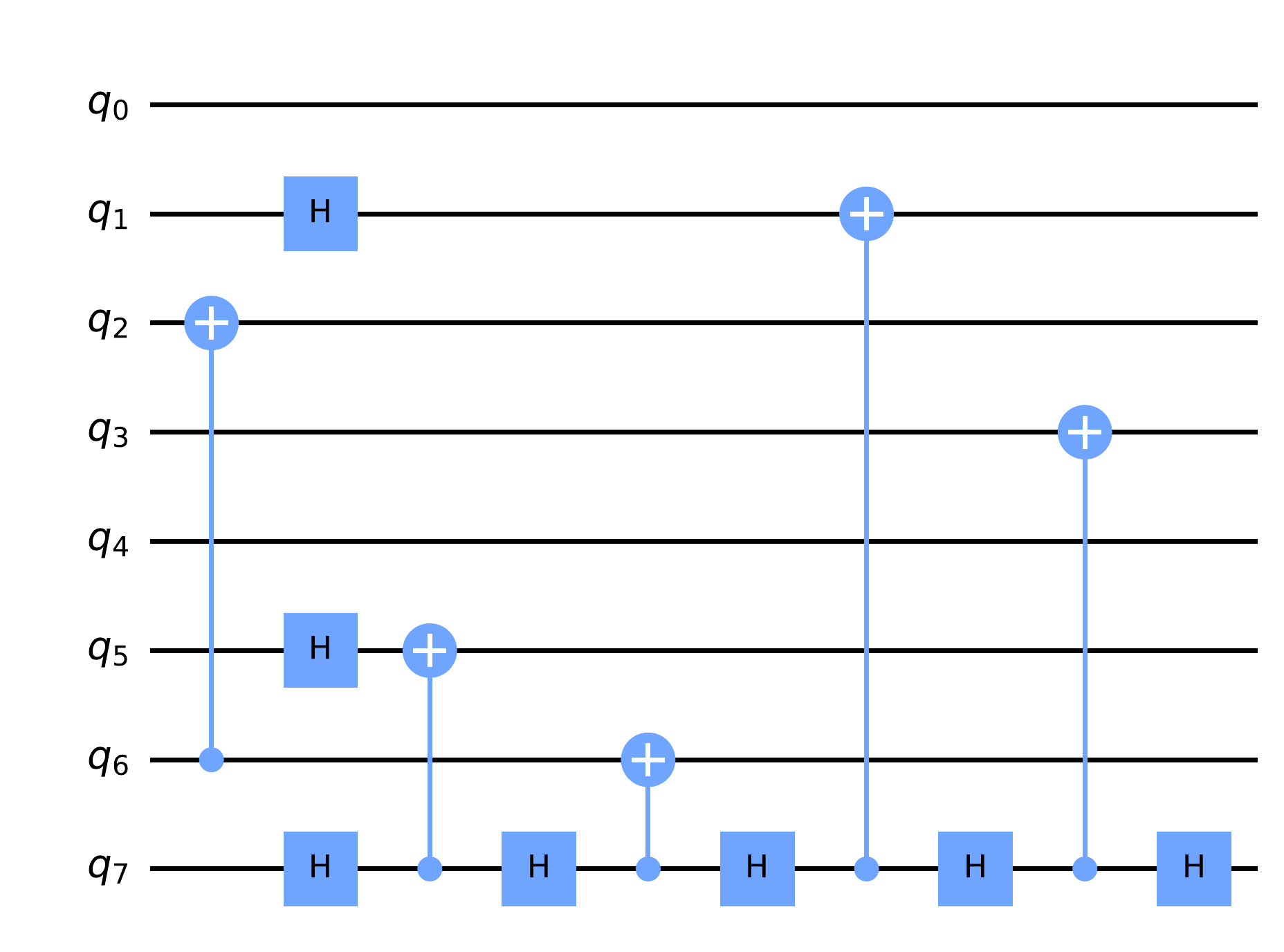
     }}
     ~
     \subfloat[\label{fig:qlinear}]{
     \includegraphics[width=0.5\textwidth]{
     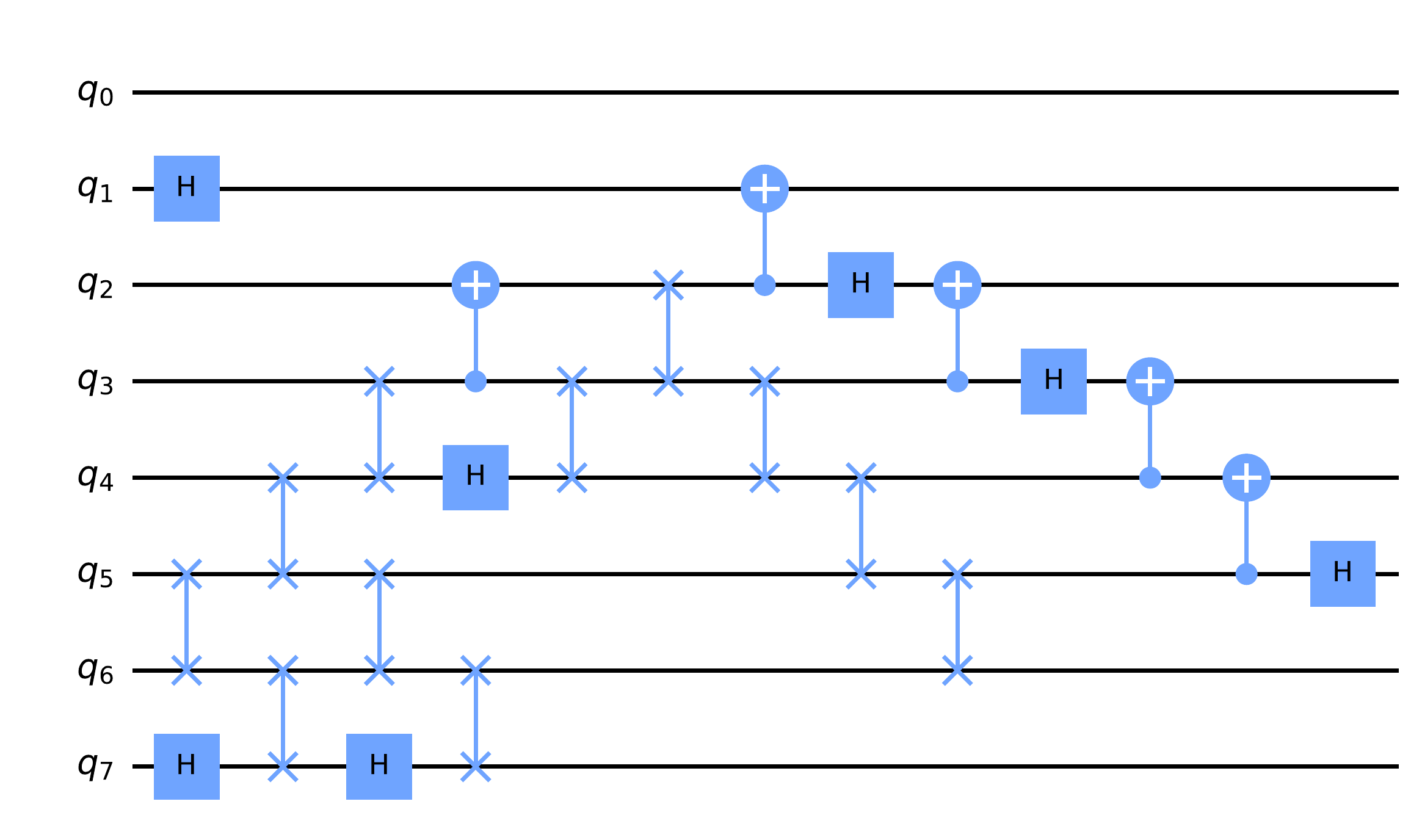
     }}
    \caption{
    Diagonalization circuits
        of a large commuting sub-Hamiltonian of a H$_4$ chain, derived in Fig.~15 of Ref.~\cite{Miller:2022sol}.
        We assume a device~\protect\subref{fig:qfull} with full qubit connectivity and~\protect\subref{fig:qlinear} with linear qubit connectivity.
        In the latter case, we apply the methods of Sec.~\ref{sec:qconnect}, involving extra {\small SWAP} gates in order to implement more local two-qubit operations. \cyan{We use the no optimization strategy to find a null space vector throughout the diagonalization.}
        \label{fig:connectivities}
    }
\end{figure*}

The nine remaining sets of commuting Pauli operators are much simpler than the one discussed above.
In the previous literature, the authors were able to derive simple circuits, notably with no SWAP gates and at most two two-qubit gates.
Our algorithm produces similar results diagonalizing the Pauli operators in each set with no {\small SWAP} gates and at most two two-qubit gates per set.

\section{Conclusions}
\label{sec:conclude}

This paper introduces an algorithm to simultaneously diagonalize Pauli operators, whose combinations are frequently used to compute observables as well as unitary operations such as in quantum simulation algorithms.
Due to the iterative nature of the algorithm, in which we seek to diagonalize all operators over one qubit at a time, we find this algorithm is additionally convenient in adaptations to quantum hardware on which qubit connectivity is very limited.
Moreover, we are able to see from analytic arguments that the scaling for depths of the resulting circuits prescribed by our algorithm is exceptionally low, scaling linearly in the total number of qubits and logarithmically in the number of independent generators of the operators.
In kind, we can analytically count the total number of two-qubit gates to be competitive with other recent constructive diagonalization algorithms.

In this paper, we have further demonstrated that the circuits we produce maintain short circuit depth and low two-qubit gate count in example Hamiltonians such as random Hermitian operators and those describing molecular systems.
It remains to be demonstrated whether the circuit depth obtained here can be substantially improved or is in fact optimal.
Likewise, we have seen that there are choices in our algorithm that can be exploited to further accommodate limited connectivity, though making the optimal such choice may be formulated as a generalized TSP.
Therefore, further work can be done to optimize these choices based upon the particular hardware in consideration.
Nevertheless, we expect that the framework for diagonalization suggested by our algorithm will enable future work to accommodate such hardware.

Additionally, let us point out that algorithms that form the clusters of commuting operators to be diagonalized must be designed in parallel with these diagonalization algorithms to produce the simplest circuits possible overall.
As such, developments in forming commuting clusters of operators would similarly assist in the effort to efficiently diagonalize such clusters overall as well.

\begin{acknowledgments}
We thank {\ora Ophelia Crawford, Ridwan Syed, Ewout van den Berg, and Daochen Wang} for helpful correspondence.
This material is based upon work supported in part by the U.S.~Department of Energy, Office of Nuclear Physics under Grants No.~DE-SC0021143 and DE-FG02-95ER40907.
The modules {\small QISKIT}~\cite{Qiskit}, {\small NUMPY}~\cite{harris2020array}, {\small GALOIS}~\cite{Hostetter_Galois_2020}, and {\small NETWORKX}~\cite{networkx_2008} have been significantly used throughout this project.
\end{acknowledgments}

\bibliography{references}

\end{document}

%% file: diagonalization_pseudocode.tex


\begin{algorithm}[H]
\caption{Qubitwise diagonalization}\label{alg:diag}
 \hspace*{\algorithmicindent} \textbf{Input:} A set of commuting Pauli operators acting on $n$ qubits. \\
 \hspace*{\algorithmicindent} \textbf{Output:} A circuit simultaneously diagonalizing the set.  
\begin{algorithmic}[1]
\State $\alpha \gets 1$
\State Discard all the qubits on which all Paulis are already diagonal.
\State $n^{(\alpha)} \gets $ Number of remaining qubits 
\If{ $n^{(\alpha)} = 0$ }
    \State Exit \Comment{The diagonalization is complete.}
\EndIf
\State Find independent Pauli operators $T^{(\alpha)}$ \Comment{See Sec.~\ref{sec:problem}.}
\State $\mathcal{M}^{(\alpha)} \gets \Big(\mathcal{X}^{(\alpha)} \big| \mathcal{Z}^{(\alpha)}\Big)$ \Comment{Tableau encoding of $T^{(\alpha)}$}
\State Find the null space of $\mathcal{M}^{(\alpha)}$
\State Select a vector $(\boldsymbol{v},\boldsymbol{w})$ in the null space of $\mathcal{M}^{(\alpha)}$
\State Select $i$ such that $v^i = 1$ or $w^i = 1$  \Comment{qubit $i$ on which all Paulis will be diagonalized}
\For{$j = 1$ to $j = n^{(\alpha)}$}
\If{$v^j = 0$ AND $w^j = 1$}
    \State Perform a conjugation with $\mathrm{H}\left(j\right)$
\EndIf
\If{$v^j = 1$ AND $w^j = 1$}
    \State Perform a conjugation with $\mathrm{S}\left(j\right)$
    \State Perform a conjugation with $\mathrm{H}\left(j\right)$
\EndIf
\EndFor
\For{$j = 1$ to $j = n^{(\alpha)}$}
\If{($v^j = 1$ or $w^j = 1$) AND $i \neq j$}
    \State Perform a conjugation with {CNOT}$(j,i)$
\EndIf
\EndFor
\State $\alpha \gets \alpha + 1$
\State Go to Line 2.
\end{algorithmic}
\end{algorithm}



%% file: modified_step2_pseudocode.tex
\begin{algorithm}[H]
\caption{Step 2 yielding logarithmic depth 
}\label{alg:diag2}
 \hspace*{\algorithmicindent} \textbf{Input:} A null space vector $(\boldsymbol{v}, \boldsymbol{w})$. \\
 \hspace*{\algorithmicindent} \textbf{Output:} A modified step (2) of the algorithm.  
\begin{algorithmic}[1]

\State{$Q \leftarrow \{q=1,\ldots, n^{(\alpha)} \mid v^{q} = 1$ or $w^{q} = 1\}$}

\While{$| Q | > 1$}
    \For{$j=1$ to $j = \lfloor |Q|/2\rfloor$}
    \State Perform a conjugation with {CNOT}$(q_{2j},q_{2j-1})$
    \EndFor
    \State{$Q \leftarrow Q\setminus\{q_{2j}\mid j=1,\ldots,\lfloor |Q|/2\rfloor\}$ }
\EndWhile
\end{algorithmic}
\end{algorithm}

